\documentclass[12pt]{article}  

\RequirePackage{graphicx,color}
\usepackage{amssymb,amsmath,slashed,mathrsfs,cite}

\usepackage{bbold,parskip,subcaption}
\usepackage{mathrsfs}
\usepackage{amsmath}
\usepackage{hyperref}
\usepackage{xcolor}

\usepackage{graphicx}% Include figure files
\usepackage{dcolumn}% Align table columns on decimal point
\usepackage{bm}% bold math
\DeclareMathAlphabet{\mathsfit}{T1}{\sfdefault}{\mddefault}{\sldefault}
\SetMathAlphabet{\mathsfit}{bold}{T1}{\sfdefault}{\bfdefault}{\sldefault}

\makeatletter
\newsavebox{\@brx}
\newcommand{\llangle}[1][]{\savebox{\@brx}{\(\m@th{#1\langle}\)}%
  \mathopen{\copy\@brx\mkern2mu\kern-0.9\wd\@brx\usebox{\@brx}}}
\newcommand{\vrt}[1][]{\savebox{\@brx}{\(\m@th{#1\vert}\)}%
  \mathopen{\copy\@brx\mkern2mu\kern-0.9\wd\@brx\usebox{\@brx}}}  
\newcommand{\rrangle}[1][]{\savebox{\@brx}{\(\m@th{#1\rangle}\)}%
  \mathclose{\copy\@brx\mkern2mu\kern-0.9\wd\@brx\usebox{\@brx}}}
\makeatother

\def\0{\mbox{\boldmath$\displaystyle\boldsymbol{0}$}}

\def\x{\mbox{\boldmath$\displaystyle\boldsymbol{x}$}}

\begin{document}

\date{\empty}

%\noindent
%\textbf{\Large The quantum theory of the spin-half mass dimension one fields}
%\\
%\\
%\textbf{\Large The quantum field theory of the spin-half bosons and fermions with mass dimension one}
%\noindent
%\textbf{\Large Irreducible unitary representations of extended Lorentz group with Wigner degeneracy }

\noindent
%\textbf{\large Irreducible unitary representations of the extended Poincar\'e group with a two-fold Wigner degeneracy: Dark matter}

\noindent
%\textbf{\large Beyond sterile neutrinos and Pontecorvo framework}
\textbf{Significance of classically forbidden regions for short baseline neutrino experiments}

%\textbf{\large Implications of a neV neutrino mass eigenstate for short baseline experiments}

%\textbf{\large LSND-KARMEN anomaly and JSNS$^2$}

%and non-vanishing amplitudes for space-like source-detector separations }
\vspace{11pt}

\noindent
{\textbf{Dharam Vir Ahluwalia\footnote{The author declares that during the LSND experiment he was at the Physics Division of the Los Alamos National Laboratory as a Director's Postdoctoral Fellow.  
} \\
}
\vspace{-21pt}
\begin{quote}
\small{
\noindent 
Center for the Studies of the Glass Bead Game\\ 220 Normanby Road, Notting Hill,  Victoria 3168, Australia }  \\ 
Email: dharam.v.ahluwalia@gmail.com 
\end{quote}
\vspace{5pt}
\textcolor{red}{\hrule}
\vspace{5pt}
\begin{quote}

\end{quote}

\vspace{5pt}
\begin{quote}
\textbf{Abstract.}
Classically forbidden regions ($\mathtt{CFRs}$) are common to both non-relativistic quantum mechanics, and to relativistic quantum field theory. It is known since 2001 that 
$\mathtt{CFR}$ contributes roughly sixteen percent of energy to the ground state of a simple harmonic oscillator
(\textsc{Adunas G. Z.} \emph{et al.}, \emph{Gen. Relativ. Gravit.}, \textbf{33} (2001) 183).
 Similarly, quantum field theoretic arguments yield a non-zero amplitude for a massive particle to cross the light cone (that is, into the $\mathtt{CFR}$). The signs of these amplitudes are opposite for fermions  and antifermions. This has given rise to an erroneous conclusion that amplitude to cross the lightcone is identically zero. This is true as long as a measurement does not reveal the considered object to be a particle or antiparticle. However, neutrino oscillation experiments do measure a neutrino $\nu$, or an antineutrino $\bar\nu$. Here we show that in the context of neutrino oscillations these observations have the potential to resolve various short baseline anomalies for a sufficiently light lowest mass eigenstate. In addition, we make a concrete prediction for the upcoming results to be announced later this year by JSNS$^2$.

\noindent
\textbf{Journal reference.} Europhysics Letters (EPL) 142 (2023) 22001.

\vspace{11pt}

\noindent
\textbf{Keywords.}
 Classically forbidden regions, short baseline neutrino anomalies,
 LSND, KARMEN, JSNS$^2$

 %Michael Peskin and Daniel Schroeder argue to the contrary. 
\end{quote}
\newpage
\noindent
%\textbf{Heuristics.}
\textbf{Introduction.}

The 1996 claim by LSND collaboration for $\overline{\nu}_\mu \to \overline\nu_e$ oscillation, and later of $\nu_\mu\to\nu_e$ oscillation,  remains unconfirmed by the KARMEN neutrino oscillation experiment~\cite{LSND:1996ubh,LSND:1997vun,Eitel:2000ry,Yellin:1998xb}. The physics community is in general agreement with Yellin, that ``both experiments have competent personnel''  and ``both have been working a long enough time to eliminate serious mistakes''~\cite{Yellin:1998xb}. Concurrently, similar anomalies exist in the short baseline experiments with reactors~\cite{Antonelli:2020uui,Giunti:2021iti}, and just a few months ago
BEST has further re-confirmed the gallium anomaly~\cite{Barinov:2021asz}.
Under these circumstances 
it is natural to suspect that all these short baseline anomalies cannot be entirely understood within the Pontecorvo framework without invoking a family of sterile neutrinos, or to look for a physical origin that has somehow evaded physicists.  

It may be relevant to parenthetically note that in the context of gallium anomaly
Giunti et al. also conclude that,
``one should pursue other possible solutions beyond short-baseline oscillations'' and further ``that the neutrino oscillation explanation of the Gallium Anomaly is in strong tension with the solar bound on active-sterile neutrino mixing'' \cite{Giunti:2022btk}.

\noindent
\textbf{Heuristics and their limitations}

To initiate a possible first-principle solution beyond short-baseline neutrino oscillations, 
it is expedient to  start with Steven Weinberg's observation in~\cite{Weinberg:1972kfs}:
`` \ldots The uncertainty principle tells us that when a particle is at position $\x_1$ at time $t_1$, we cannot also define its velocity precisely. In consequence there is certain chance of a particle getting from $x_1$ to $x_2$ even if $x_1-x_2$ is spacelike, that is $\vert \x_1-\x_2\vert > \vert {x_1}^0 - {x_2}^0\vert$. To be more precise, the probability of a particle reaching $x_2$ if it starts at $x_1$ is nonnegligible as long as
\begin{align}
(\x_1 -\x_2)^2 - ({x_1}^0 -{x_2}^0)^2 \le \frac{\hbar^2}{m^2}
\end{align}
where $\hbar$ is Planck's constant (divided by $2\pi$) and $m$ is the particle mass. (Such spacetime intervals are very small even for elementary particle masses; for instance, if $m$ is  the mass of a proton then ${\hbar}/{m} = 2\times 10^{-14}$ cm \ldots)''

Tony Zee has also noted that ``classically, a particle cannot get outside the lightcone, but a quantum field can ``leak'' out over a distance of order $\hbar/m$ by the Heisenberg uncertainty principle''~\cite[p. 24]{Zee:2003mt}. \thefootnote{ Note: Here we have inserted $\hbar$ in order that both quotes are in the same units.     }
%\end{quote}

Since only the mass squared differences are observable in the Pontecorvo framework the mass of the lowest mass eigenstate remains free (modulo cosmological and astrophysical constraints~\cite{Wang:2017htc,Lattanzi:2017ubx,TopicalConvenersKNAbazajianJECarlstromATLee:2013bxd}). For the sake of our argument, let its mass be 
of the order of a nano electron volt
$(\mbox{neV})$,\footnote{This may appear a bit audacious. But to the best of our knowledge such small mass scale is not prohibited by any existing data. However, once such an assumption is made some very concrete predictions follow.}
 that is roughy eighteen orders of magnitude smaller than a proton's  mass. Then, ${\hbar}/{m} \sim  100~ \mbox{meters}$. 
This covers the source-detector distance for all the mentioned short baseline experiments.

However, these heuristics can be misleading as in the $m\to 0$ limit the light cone completely opens up! As such one must seek a fully quantum field theoretic expression for ``leaking off'' into the $\mathtt{CFR}$ that does not suffer from this problem.

\noindent
 \textbf{Beyond the heuristics.}
 
To go beyond the heuristics, in quantum field theoretic
framework if $x$ and $x^\prime$ are separated by a space-like interval then the amplitude for a fermionic \emph{particle}
to reach $x^\prime$ starting from $x$ is~\cite[eq. 2.13]{Ahluwalia:2011rg}
\begin{align}
A(x\to x^\prime)  =  +\,\frac{1}{\pi^2}
\frac{m^2}{\sqrt{r^2-t^2}} K_1\big(m\sqrt{r^2-t^2}\big),
\label{eq:a}
\end{align}
where $r=\vert \x^\prime-\x\vert$, $t=\vert x^{\prime 0} - x^0 \vert$, and   $K_\nu (\ldots)$ is the modified Bessel function of the second kind of order $\nu$. % We use natural units: $\hbar=c=1$.
The total amplitude contributing at $r$ from \emph{all} spacelike separations -- that is, from the entire $\mathtt{CFR}$ --  is calculated by integrating $A(x\to x^\prime) $ from $t=0$ to $t=r$. 
To the leading order (that is, approximating $A(x\to x^\prime)$  to 
$\mathcal{O}(t^2)$ in the series expansion, and implementing the indicated integration)
\begin{equation}
A (r,\mathtt{CFR})= +\,
\frac{m^2}{6 \pi^2 r} \big(  m r^2 K_2\left(  m r\right)+ 6 r K_1\left(m r\right)
\big).
\label{eq:acfr}
\end{equation}
For \emph{antiparticles} the plus sign in equations (\ref{eq:a}) and  (\ref{eq:acfr}) after the equal sign should be changed to a minus sign.
There is also a multiplicative factor $\eta$ on the right hand sides of (\ref{eq:a}) and (\ref{eq:acfr}) that is
determined from the requirement that the total probability of a particle being in the $\mathtt{CAR}$ plus the $\mathtt{CFR}$ must be unity
\begin{align}
\eta \stackrel{\mathrm{def}}{=} \sqrt{\eta^\ast \eta} = \left(\frac{1}{A^\ast A \vert_{\mathtt{CAR}} + A^\ast A \vert_{\mathtt{CFR}}}\right)^{1/2}.
\end{align}
Since the probability for $\mathtt{CAR}$ ($\mathtt{CAR}$, is abbreviation for classically allowed region)  is $\gg$ 
probability for $\mathtt{CFR}$,   the $\eta$ upto a possible phase factor may be taken as
\begin{equation}
\eta \approx \frac{1}{\sqrt{A^\ast A}\vert_{\mathtt{CAR}}}
\end{equation}

\begin{figure}
\includegraphics{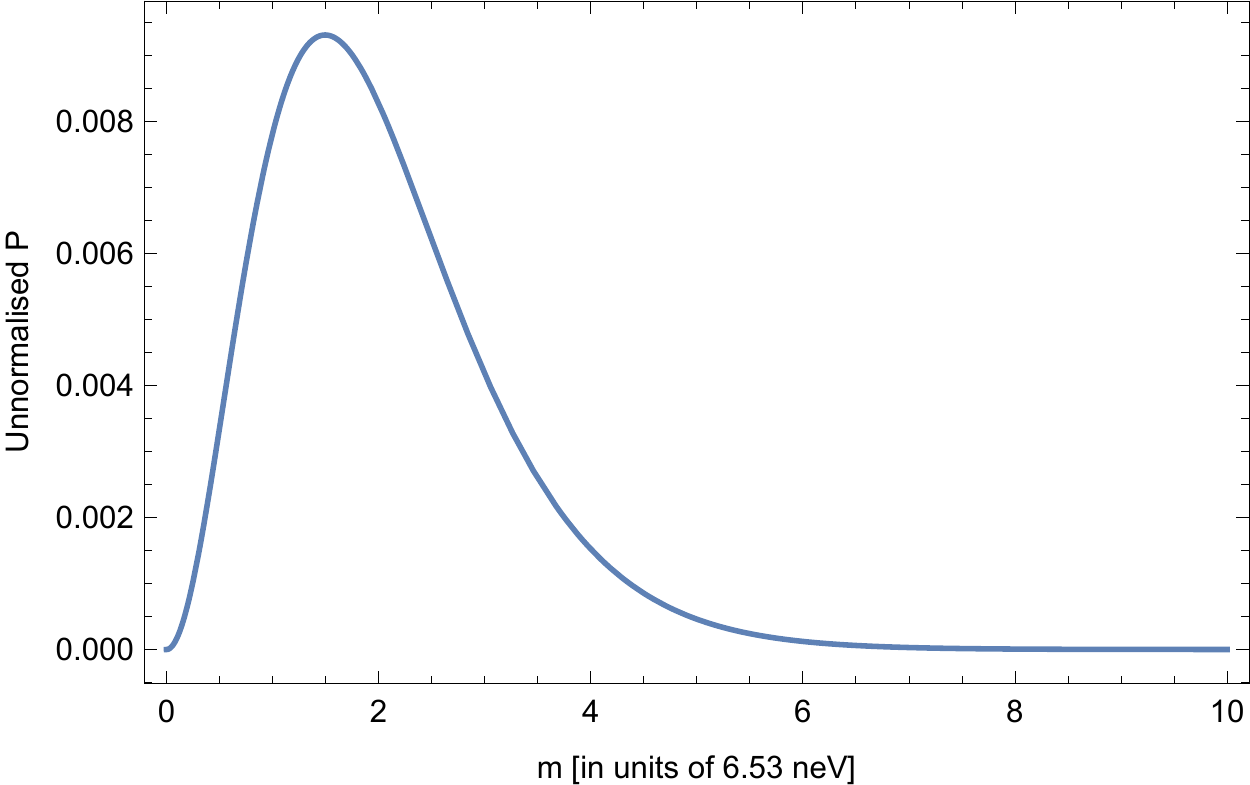}
\caption{Unnormalised $\mathcal{P}(x\to x^\prime) $ for a fixed $r=1 \,(\mbox{in units}\, \mbox{6.53 neV})^{-1}$.}
\end{figure}

As such $\eta$ is practically constant in the $\mathtt{CFR}$ and is not explicitly noted in our considerations.
To illustrate the $m$ dependence of the probability 
$\mathcal{P}(x\to x^\prime) $
that derives from (\ref{eq:acfr}) we plot the un-normalised $\mathcal{P}$ as a function of $m$.
This is depicted in Figure 1.
The result deviates somewhat from what would be expected from the heuristics: for $m= 0$ the probability to go off the light-cone is zero. For a given $r$, in $\mathtt{CFR}$ it increases to a maximum and then falls off and vanishes again for $m\to\infty$.

%\noindent
%\textbf{Return to heuristics}

An intuitive way to understand the inevitability of our results is to realise that almost all quantum systems are endowed with $\mathtt{CFRs}$. Quantum tunneling is just one such example in which one may access $\mathtt{CFRs}$  . Explicit calculations for a simple harmonic oscillator in its ground state are given in reference~\cite{Adunas:2000zn}; where it was found that $\mathtt{CFR}$'s contribution to the ground state energy equals $\left[1-\mbox{erf}(1)\right] \times \frac12\hbar \omega$.\footnote{The notation is standard, and $\mbox{erf}(...)$ denotes the error function.} Since $\left[1-\mbox{erf}(1)\right] \approx 0.16$, 
roughly $16\%$ of the ground state energy is contributed by the $\mathtt{CFR}$.
If one considers a quantum field as a system of infinitely many simple harmonic oscillators, then that a massive particle may have a finite probability to access $\mathtt{CFR}$ becomes transparent.

\noindent 
\textbf{Significance for short-baseline neutrino experiments}

For comparing LSND with KARMEN  we take units such that $r=1$ for LSND.\footnote{The mass $m$ is then measured in units of $6.53$ neV.} Then $r=17.6/30$ for KARMEN. The ratio of the number of events expected for LSND versus KARMEN for the $\bar\nu_\mu\to \bar\nu_e$ channel is then
\begin{equation}
\alpha \,\frac{\mathcal{P}(x\to x^\prime) \big\vert_{r=1}}
{\mathcal{P}(x\to x^\prime) \big\vert_{r=17.6/30}}\label{eq:LK}
\end{equation}
with $\alpha$ introduced to account for various detector/beam-related parameters
\begin{equation}
\alpha \stackrel{\mathrm{def}}{=} \left(\frac{20000\, \mbox{Coulombs}}{ 2897\,\mbox{Coulombs}}\right) 
\left( \frac{167 \,\mbox{tons}}{56\, \mbox{tons}} \right)
\left(\frac{17.6\, \mbox{meters}}{30.0 \,\mbox{meters}}\right)^2
\end{equation}

\begin{figure}
\includegraphics{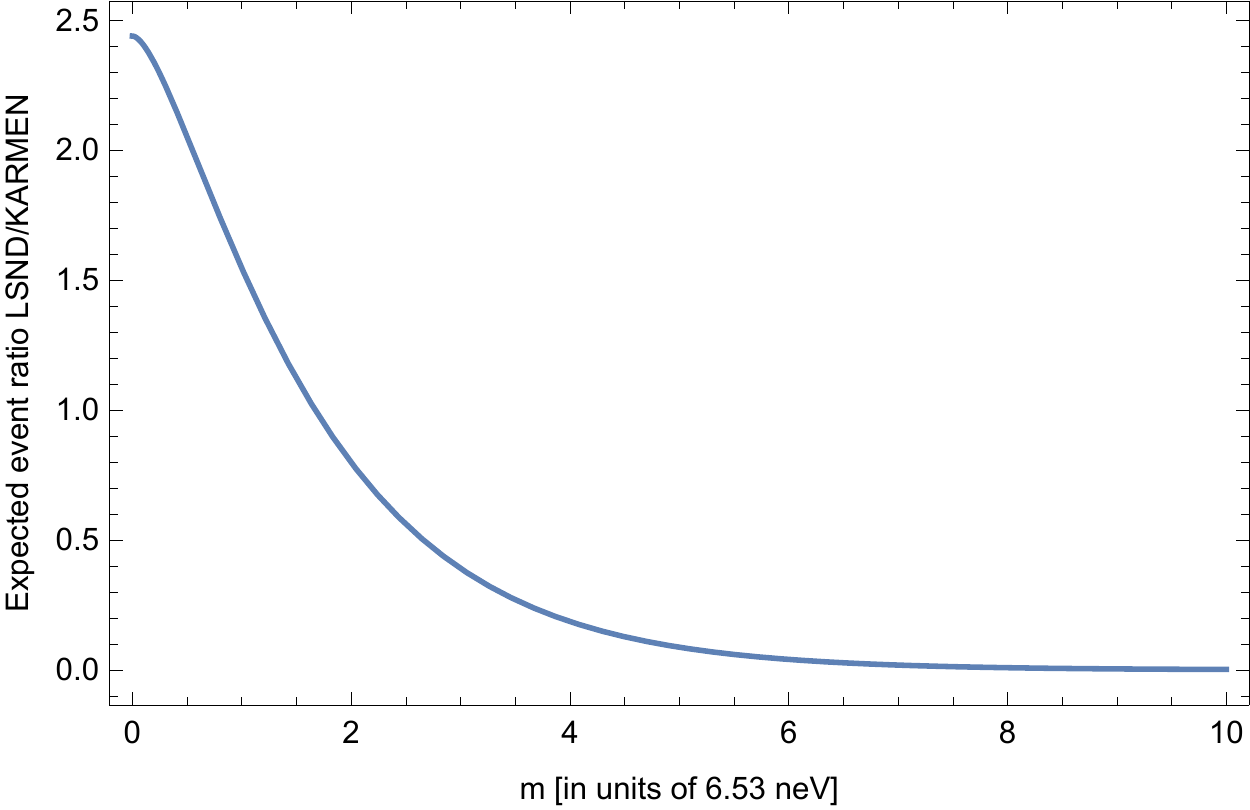}
\caption{The 
ratio of the number of events expected for LSND versus KARMEN as a function of $m$ in the $\bar\nu_\mu\to \bar\nu_e$ mode.
For the $\nu_\mu\to \nu_e$ mode, LSND gains an additional advantage by a factor $34$ due to target differences and drift spaces -- see~\cite{Yellin:1998xb} for details.
}
\end{figure}

The first term in $\alpha$ represents  data 
for KARMEN (three months after the 1996 upgrade):  
$2897$ coulombs of integrated proton beam;  for LSND  with over 20000 coulombs collected between 1993 and 1997. 
The second term accounts for the difference in the detector volumes. The third term accounts for source-detector distance: $30\,\mbox{meters}$ for LSND versus $17.6\, \mbox{meters}$ for KARMEN; it accounts for diminution in the flux.
Other issues, such as different duty factors, can apparently be incorporated (though this author is not competent enough to do that). Our source for these details is reference~\cite{Yellin:1998xb}.

Figure 2 depicts the 
ratio of the number of events expected for LSND versus KARMEN as a function of $m$. In the stated  context,
there is an indication in favour of LSND. 

In making these estimates we have assumed that the dominant contribution comes from the lowest lying mass eigenstate. The mixing matrix element that enters in the projection to $\bar\nu_e$ from the lowest mass eigenstate cancels in taking the ratio~(\ref{eq:LK}). For reactor experiments one must be careful: while for the short baseline experiments, ``leaking'' off the lightcone may be important; the Pontecorvo mechanism may takeover for larger baselines. 

The bottom line is thus: repeat a LSND-type experiment with, if possible, KARMEN-like feature that allows for better space-time resolution of events. Such an effort is already underway in Korea's JSNS$^2$ experiment~\cite{Maruyama:2022juu}. In its first phase  the source-detector distance is $24$ meters. In its second phase an additional detector will be placed at a distance of $48$ meters.
It is designed to be a direct test of the $\overline\nu_e$ excess events observed by LSND. With $m$ measured in units of $6.53$ neV, the ratio of the number of events per unit volume of the far versus the near detector as a function of $m$ reads
\begin{align}
\frac{\left(15 K_1\left(\frac{8 m}{5}\right)+4 m
   K_2\left(\frac{8 m}{5}\right)\right)^2}{4
   \left(15 K_1\left(\frac{4 m}{5}\right)+2 m
   K_2\left(\frac{4 m}{5}\right)\right)^2}
   \label{eq:JSNS2}
\end{align}
where we have included the effect of flux reduction
for the far detector relative to that of the near detector. This variation with $m$ is displayed in Figure 3.

While here our focus has been on accelerator based experiments, we take note that
in the context of reactor experiments, data taken by Neutrino-4  needs to be re-interpreted along the lines outlined here~\cite{Serebrov:2021ndf}.
The concerns about its energy resolution pointed out by Giunti et al. in ~\cite{Giunti:2021iti} then completely evaporate. Instead, because of the segmented detector design its spatio-temporal resolution becomes an advantage. Additionally, its ability to change the detector-source distance makes the Neutrino-4 an important   tool for studying the framework described here.

\noindent
\textbf{Conclusion.}

Here we have argued that LSND and KAMEN anomaly, and various other short baseline anomalies,  may be resolved by first principle arguments. While  Pontecorvo framework for neutrino oscillations allows us to understand, not only what was once called a `solar neutrino anomaly,' but also the atmospheric neutrino data, it fails to account for  various short baseline observations. 
In the process it seems that a fundamental mechanism has been missed. That mechanism is for a massive particle, with neV range mass, to cross the light-cone and trigger a detector at space-like separation.  A positive result shall not only resolve the short baseline  anomalies but it shall also provide a mass of the lowest mass eigenstate without invoking a sterile neutrino.

\begin{figure}
\includegraphics{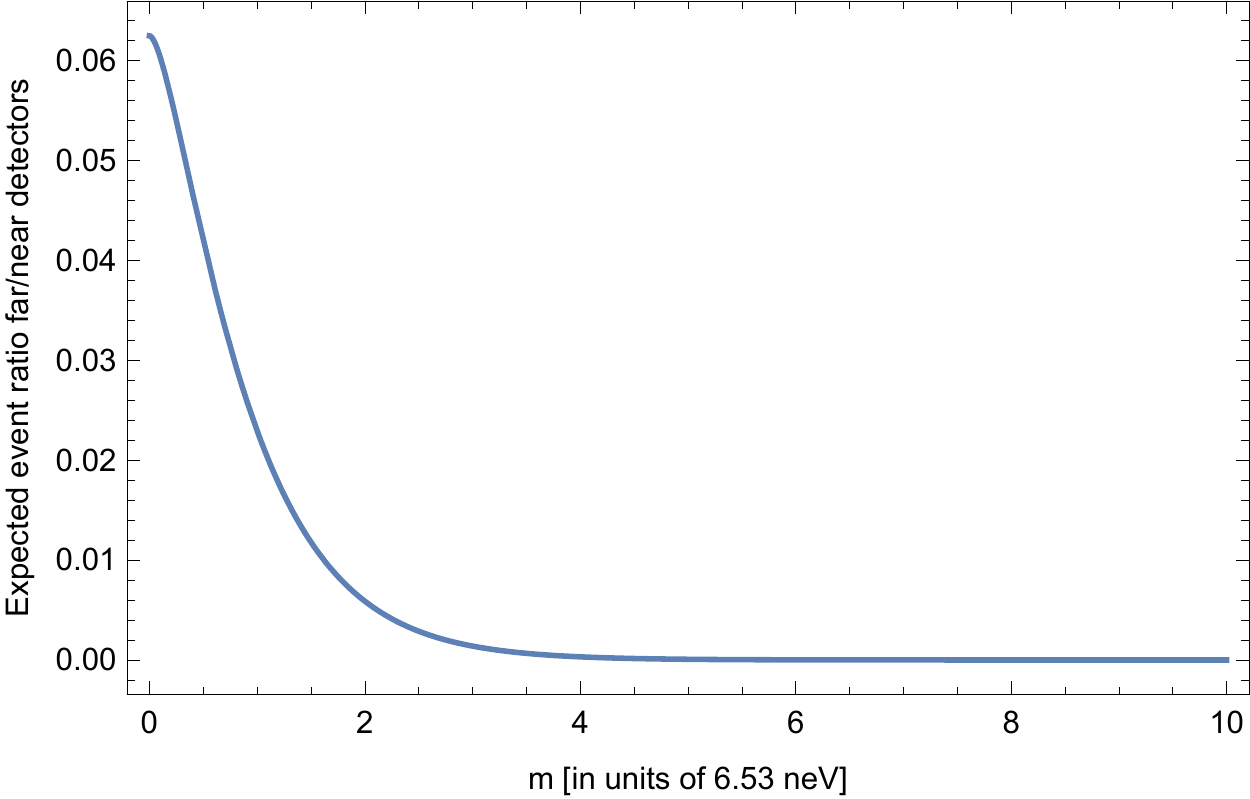}
\caption{The ratio given by equation (\ref{eq:JSNS2}) as a function of $m$.}
\end{figure}
\noindent
\textbf{Acknowledgements}

We thank Sebastian Horvath and Dimitri Schritt for our earlier collaboration on related ideas~\cite{Ahluwalia:2011rg}, and Cheng-Yang Lee and Julio Hoff da Silva  for  their helpful comments on the  draft of this Letter. It is also our pleasure to thank Bill Louis for his suggestions, and for bringing JSNS$^2$ to our attention.

%\bibliographystyle{JHEP-2}       
% included to cite titles of references, and should not be edited out
%\bibliography{CFR-in-QFT-bib}   % name your BibTeX data base

\begin{thebibliography}{10}

\bibitem{LSND:1996ubh}
{\bf LSND} Collaboration, C.~Athanassopoulos {\em et.~al.}, {\it {Evidence for
  $\overline{\nu}_\mu\to\overline{\nu}_e$ neutrino oscillations from the LSND
  experiment at LAMPF}},  {\em Phys. Rev. Lett.} {\bf 77} (1996) 3082--3085
  [\href{http://arXiv.org/abs/nucl-ex/9605003}{{\tt nucl-ex/9605003}}].

\bibitem{LSND:1997vun}
{\bf LSND} Collaboration, C.~Athanassopoulos {\em et.~al.}, {\it {Evidence for
  $\nu_\mu \to \nu_e$ neutrino oscillations from LSND}},  {\em Phys. Rev.
  Lett.} {\bf 81} (1998) 1774--1777
  [\href{http://arXiv.org/abs/nucl-ex/9709006}{{\tt nucl-ex/9709006}}].

\bibitem{Eitel:2000ry}
K.~Eitel, {\it {Statistical analysis of the LSND evidence and the KARMEN
  exclusion for $\overline{\nu}_\mu\to\overline{\nu}_e$ oscillations}},  in
  {\em {Workshop on Confidence Limits}}, pp.~173--185, 8, 2000.

\bibitem{Yellin:1998xb}
S.~J. Yellin, {\it {A Comparison of the LSND and KARMEN anti-neutrino
  oscillation experiments}},  {\em AIP Conf. Proc.} {\bf 478} (1999), no.~1
  416--421 [\href{http://arXiv.org/abs/hep-ex/9902012}{{\tt hep-ex/9902012}}].

\bibitem{Antonelli:2020uui}
V.~Antonelli, L.~Miramonti and G.~Ranucci, {\it {Present and Future
  Contributions of Reactor Experiments to Mass Ordering and Neutrino
  Oscillation Studies}},  {\em Universe} {\bf 6} (2020), no.~4 52.

\bibitem{Giunti:2021iti}
C.~Giunti, Y.~F. Li, C.~A. Ternes and Y.~Y. Zhang, {\it {Neutrino-4 anomaly:
  oscillations or fluctuations?}},  {\em Phys. Lett. B} {\bf 816} (2021) 136214
  [\href{http://arXiv.org/abs/2101.06785}{{\tt 2101.06785}}].

\bibitem{Barinov:2021asz}
V.~V. Barinov {\em et.~al.}, {\it {Results from the Baksan Experiment on
  Sterile Transitions (BEST)}},  {\em Phys. Rev. Lett.} {\bf 128} (2022),
  no.~23 232501 [\href{http://arXiv.org/abs/2109.11482}{{\tt 2109.11482}}].

\bibitem{Giunti:2022btk}
C.~Giunti, Y.~F. Li, C.~A. Ternes, O.~Tyagi and Z.~Xin, {\it {Gallium Anomaly:
  critical view from the global picture of \ensuremath{\nu}$_{e}$ and $
  {\overline{\nu}}_e $ disappearance}},  {\em JHEP} {\bf 10} (2022) 164
  [\href{http://arXiv.org/abs/2209.00916}{{\tt 2209.00916}}].

\bibitem{Weinberg:1972kfs}
S.~Weinberg, {\em {Gravitation and Cosmology}: {Principles and Applications of
  the General Theory of Relativity}}.
\newblock John Wiley and Sons, New York, 1972.
\newblock {See Ch. 2, Sec. 13}.

\bibitem{Zee:2003mt}
A.~Zee, {\em {Quantum field theory in a nutshell}}.
\newblock {Princeton University Press}, 2003.
\newblock {Second Edition}.

\bibitem{Wang:2017htc}
S.~Wang, Y.-F. Wang and D.-M. Xia, {\it {Constraints on the sum of neutrino
  masses using cosmological data including the latest extended Baryon
  Oscillation Spectroscopic Survey DR14 quasar sample}},  {\em Chin. Phys. C}
  {\bf 42} (2018), no.~6 065103 [\href{http://arXiv.org/abs/1707.00588}{{\tt
  1707.00588}}].

\bibitem{Lattanzi:2017ubx}
M.~Lattanzi and M.~Gerbino, {\it {Status of neutrino properties and future
  prospects - Cosmological and astrophysical constraints}},  {\em Front. in
  Phys.} {\bf 5} (2018) 70 [\href{http://arXiv.org/abs/1712.07109}{{\tt
  1712.07109}}].

\bibitem{TopicalConvenersKNAbazajianJECarlstromATLee:2013bxd}
K.~N. Abazajian {\em et.~al.}, {\it {Neutrino Physics from the Cosmic Microwave
  Background and Large Scale Structure}},  {\em Astropart. Phys.} {\bf 63}
  (2015) 66--80 [\href{http://arXiv.org/abs/1309.5383}{{\tt 1309.5383}}].

\bibitem{Ahluwalia:2011rg}
D.~V. Ahluwalia, S.~P. Horvath and D.~Schritt, {\it {Amplitudes for space-like
  separations and causality}},  \href{http://arXiv.org/abs/1110.1162}{{\tt
  1110.1162}}.

\bibitem{Adunas:2000zn}
G.~Z. Adunas, E.~Rodriguez-Milla and D.~V. Ahluwalia, {\it {Probing quantum
  violations of the equivalence principle}},  {\em Gen. Rel. Grav.} {\bf 33}
  (2001) 183--194 [\href{http://arXiv.org/abs/gr-qc/0006022}{{\tt
  gr-qc/0006022}}].

\bibitem{Maruyama:2022juu}
T.~Maruyama, {\it {The status of JSNS$^2$ and JSNS$^2$-II}},  {\em PoS} {\bf
  NuFact2021} (2022) 159.

\bibitem{Serebrov:2021ndf}
A.~P. Serebrov, R.~M. Samoilov and M.~E. Chaikovskii, {\it {Analysis of the
  result of the Neutrino-4 experiment in conjunction with other experiments on
  the search for sterile neutrinos within the framework of the 3 + 1 neutrino
  model}},  \href{http://arXiv.org/abs/2112.14856}{{\tt 2112.14856}}.

\end{thebibliography}

\providecommand{\href}[2]{#2}\begingroup\raggedright\endgroup
\end{document}